\documentclass[prd,twocolumn,aps,amsmath,nofootinbib,superscriptaddress]
{revtex4}

\usepackage{graphicx}
\usepackage{amsfonts}
\usepackage[usenames]{color}
\usepackage{amsmath}

\def\Z{\mathbb{Z}}
\def\R{\mathbb{R}}
\def\C{\mathbb{C}}
\def\Q{\mathbb{Q}}

\def\SO{\mathop{\rm SO}}

\def\SU{\mathop{\rm SU}}

\def\simgt{\mathrel{\lower2.5pt\vbox{\lineskip=0pt\baselineskip=0pt
           \hbox{$>$}\hbox{$\sim$}}}}
\def\simlt{\mathrel{\lower2.5pt\vbox{\lineskip=0pt\baselineskip=0pt
           \hbox{$<$}\hbox{$\sim$}}}}

\newcommand{\bea}{\begin{eqnarray}}
\newcommand{\eea}{\end{eqnarray}}

\newcommand{\comment}[1]{}

\begin{document}


\preprint{KCL MTH-14-13, IPMU14-0266}

\title{Distribution of the Number of Generations in Flux Compactifications}

\author{Andreas P. Braun}
\affiliation{Department of Mathematics, King's College, London WC2R 2LS, UK}

\author{Taizan Watari}
\affiliation{Kavli Institute for the Physics and Mathematics of the Universe, 
the University of Tokyo, Tokyo, 277-8583, Japan} 

\begin{abstract}
Flux compactification of string theory generates an ensemble with a large 
number of vacua called the landscape. By using the statistics 
of various properties of low-energy effective theories in the string landscape, 
one can therefore hope to provide a scientific foundation to the notion of naturalness. 
This article discusses how to answer such questions 
of practical interest by using flux compactification of F-theory. 
It is found that the distribution is approximately in a factorized form 
given by the distribution of the choice of 7-brane gauge group, that of
the number of generations $N_{\rm gen}$ and that of effective coupling constants. 
The distribution of $N_{\rm gen}$ is approximately Gaussian 
for the range $|N_{\rm gen}| \lesssim 10$. The statistical cost of higher-rank 
gauge groups is also discussed.
\end{abstract}

\maketitle


%
\noindent
{\bf 1. {\em Introduction}:}

String theory with compactified extra dimensions gives rise to 
a large number of vacua. The diversity of vacua originates from the 
choice of topology of compact internal space and flux 
configurations on it \cite{flux}. String theory as understood 
in this way today therefore does not predict a unique low-energy effective 
theory. Despite this lack of prediction, there are still many ways in which 
we can take advantage of such an ensemble of string vacua for a 
better understanding of particle physics in the real world. Such an ensemble 
containing a large number of vacua provided by a fundamental theory is referred to as a 
landscape of vacua, or landscape for short. 

At least two ideas have been proposed so far in how to take advantage 
of such landscape of string theory. One is in the context 
of understanding the non-vanishing (yet extremely small) dark energy.
It has been realized that the very small value was predicted to be 
``natural'' under the combination of three ansatzes, 
i) that the universe is occupied with many distinct areas facilitating 
effective theories with different values of the cosmological 
constant \cite{Hawking}, 
ii) that the statistical distribution of dark energy is taken as a principle 
of arguing naturalness, and 
iii) that the observational factor (anthropics) 
is taken into account in the statistical distribution \cite{CC-1}.
The string landscape indeed give rise to such an ensemble of low-energy 
effective theories with different values of the cosmological 
constant \cite{BP}, and eternal inflation has this ensemble of 
string vacua realized in the universe \cite{Guth-eternal}.
Thus the string landscape provides a theoretical foundation for the 
attempt of understanding dark energy along the lines of i) ii) and iii).  

The string landscape can also provide a scientific foundation for a notion 
of naturalness for various kinds of parameters of the standard model, 
not just for the value of dark energy. Naturalness has been exploited 
for decades as a guiding principle in the quest of models beyond the 
standard model. Arguments relying on naturalness, however, tend 
to depend 
on the class of vacua (an ensemble) one has in mind. 
Since string theory is able to provide a well-motivated ensemble of vacua 
on which naturalness arguments can be built, this is hence another place 
where string theory can contribute to progress of theoretical particle physics.

This article aims at making progress in the second direction above. 
It is known that flux compactification of Type IIB string theory / F-theory
stabilizes not only complex structure moduli but also the brane configuration. 
This means that both gauge groups and coupling constants of the effective 
theory are determined once a topological flux configuration is given. 
Exploiting all of the theoretically possible topological flux configurations, 
an ensemble of low-energy effective theories with various 
gauge groups and coupling constants is generated in this framework, 
in principle. To get this done in practice, however, a clever approach 
is necessary. Low-energy effective theories are usually classified 
in terms of their algebraic information (such as gauge groups, matter 
representations and presence/absence of certain types of interactions) first, 
and then in terms of topological information (such as the number of generations 
of matter fields in a given representation). Effective theories with the 
same algebraic and topological information are then specified by 
the values of coupling constants. The string landscape 
will be of some use only when the statistical distribution of low-energy 
observables are presented and studied in compatibility with such a hierarchical 
classification of effective theories.

This can be done along the line described in \cite{BKW-phys}, which is 
built on top of pioneering works \cite{ADD, Denef}. The study 
of \cite{BKW-phys}, however, used K3 $\times$ K3 compactification of F-theory, 
where analysis is a little easier, but we cannot even hope to obtain a 
semi-realistic model of low-energy physics. This article applies the method to 
more general Calabi--Yau fourfolds for F-theory compactification,  
derives answers to questions of practical interest, and exemplifies 
the potential power of the method. 

Section 2 
is devoted to a review of the method in \cite{ADD, Denef, BKW-phys} 
along with new observations in \cite{BW-H22}. The method is then applied 
to a class of compactifications that lead to semi-realistic supersymmetric 
grand unification (GUT) models in section 3. 
We derive distribution of the number of generations in flux vacua 
of SU(5) GUT models, and also study how the number of flux vacua scales 
when we require SO(10), SU(5) or no unification group on 7-branes, 
respectively.
We find that the distribution is in a factorized form to 
a good approximation, independently of topological choice of geometry.
Details and explanations omitted in this article are found in \cite{BW-H22}.

\vspace{10pt}
\noindent
{\bf 2. {\em The Method}:}
\label{sec:review-ADD-BKW}

A family of geometries over a restricted moduli space, 
$\pi: {\cal Y} \longrightarrow {\cal M}_*$, is a useful concept 
when one wants to focus on flux vacua with a given algebraic information. 
A restricted family and its moduli space, 
$\pi: {\cal Y} \longrightarrow {\cal M}_*^{A4}$ (resp. ${\cal M}_*^{D5}$), 
is specified for a topological choice of $(B_3,[S])$, where $[S]$ is 
a divisor class in $B_3$. Each member of the family, 
$\pi^{-1}(p) = Y_p$ for $p \in {\cal M}_*^{A4}$ (resp. ${\cal M}_*^{D5}$),  
is a smooth elliptically fibred Calabi--Yau fourfold 
$\pi_{Y_p}: Y_p \longrightarrow B_3$ with a section, and the discriminant 
locus of the fibration $\pi_{Y_p}$ contains an irreducible 
component in $[S]$ and the generic fibre over it is $I_5$ (resp. $I_1^*$) 
in the Kodaira classification.
Any one of such fourfolds in the family over ${\cal M}_*^{A4}$ (resp. 
${\cal M}_*^{D5}$) can be used for F-theory compactification that results 
in a vacuum with an $R = \SU(5)$ (resp. $R = \SO(10)$) unification group on 
7-branes.
${\cal M}_*^{A4}$ or ${\cal M}_*^{D5}$ parametrizes the complex structure of 
such geometries.\footnote{
As in \cite{BKW-phys} and literatures therein, we only consider
flux vacua for $(B_3, [S])$ of a given topology and for a given choice of 
K\"{a}hler form $J$ on $Y_p$ that is in $H^2(Y_p; \Q) \cap H^{1,1}(Y_p; \R)$
modulo multiplication of $\R^\times$. The (restricted) moduli space 
${\cal M}$ or ${\cal M}_*$ therefore refers only to that of complex 
structure. It is beyond the scope of this article or \cite{BW-H22} 
to include the scanning over $(B_3, [S])$'s of different topology or their 
K\"{a}hler moduli spaces.}
Higher rank 7-brane gauge group implies a larger number of independent 
divisors 
\begin{equation}
[\hat{C}_i] \in [H^2(Y_p; \Z) \cap H^{1,1}(Y_p; \R)], \quad 
i=1,\cdots,{\rm rank}(R), 
\end{equation}
for a generic geometry $\pi^{-1}(p) = Y_p$ in $p \in {\cal M}_*$.

An ensemble of F-theory flux vacua with a given algebraic and topological 
information is specified by a pair $(H_{\rm scan}, G_{\rm fix}^{(4)})$, where 
\begin{equation}
H_{\rm scan} \subset [H^4(Y; \Z)]_{\rm ker}, \quad 
G_{\rm fix}^{(4)} \in [H^4(Y; \mbox{\textquoteleft} \Z \mbox{\textquoteright)}]_{\rm prim};
\end{equation}
the subscript ``prim'' implies $J \wedge G = 0 \in H^6(Y;\R)$ 
(the D-term condition), and ``ker'' both $J \wedge G = 0$ and 
$i_{\hat{C}_i}^*( G) = 0 \in H^4(\hat{C}_i; \Q)$. The last condition is to 
make sure that fluxes in $H_{\rm scan}$ do not introduce gauge symmetry 
breaking (cf \cite{BW-H22}). An ensemble of 4-form flux
\begin{equation}
 \left\{ G^{(4)}_{\rm tot} = G^{(4)}_{\rm scan} + G^{(4)}_{\rm fix} \; | \; 
   G^{(4)}_{\rm scan} \in H_{\rm scan} \right\}   
\label{eq:flux-ensemble}
\end{equation}
determines an ensemble of vacua of complex structure 
through the superpotential $W \propto \int_{Y_p} \Omega_{Y_p} \wedge G^{(4)}_{\rm tot}$.

Statistics of such an ensemble of vacua can be presented as a 
distribution over the restricted moduli space ${\cal M}_*$.
The distribution was worked out analytically in a very robust 
way \cite{ADD} for the vacuum index density $d\mu_I$,
\begin{equation}
d^{2m}z \; \sum_{G^{(4)}_{\rm scan}} \delta^{2m}(DW, \overline{DW}) 
 \; {\rm det}\left[ \begin{array}{cc}
  D^2 W & \bar{D}DW \\ D \overline{DW} & \overline{D^2W} \end{array} \right]\, ,
\label{eq:index-density-def}
\end{equation}
to which each flux vacuum on ${\cal M}_*$ contributes by a delta-function 
with coefficient $\pm 1$.
Here, $m:= {\rm dim}_{\C} {\cal M}_* = h^{3,1}(Y)$, and the
$dz$'s are local holomorphic coordinates on ${\cal M}_*$. Derivatives in 
$DW$, $\overline{DW}$ etc. are with respect to the fields corresponding 
to the complex structure moduli tangent to ${\cal M}_*$.

Making a continuous approximation \cite{ADD} of the sum over flux configurations 
$\sum_{G^{(4)}_{\rm scan}}$ in \eqref{eq:index-density-def}, 
the vacuum index density is cast into the following form
\begin{equation}
 d\mu_I = \frac{(2\pi L_*)^{K/2}}{(K/2)!} \rho_I,   \quad 
 K := {\rm dim}_{\R} (H_{\rm scan} \otimes \R)\, .   
\label{eq:ADD-formula-1}
\end{equation}
Here, $L_*$ is the maximal D3-brane charge available.  
Although $\rho_I$ depends on the choice of $H_{\rm scan}$, it is given by 
\begin{equation}
c_m(T{\cal M}_* \otimes {\cal L}) = 
{\rm det}\left( - \frac{R}{2\pi i} 
                + \frac{\omega}{2\pi} {\bf 1}_{m\times m} \right) 
\label{eq:ADD-formula-2}
\end{equation}
for the K\"{a}hler form $\omega$ on ${\cal M}_*$ 
whenever $(H_{\rm scan} \otimes \R)$ contains the real primary horizontal 
subspace $H^4_{H*}(Y;\R) \subset H^4(Y; \R)$ \cite{ADD, Denef, BKW-phys}.
$H^4_{H*}(Y; \R)$ is the real part of the primary horizontal 
subspace in \cite{mirror-4fold}, 
\begin{equation}
 {\rm Span}_{\C} \left\{ \Omega_{Y_p}, D\Omega_{Y_p},D^2\Omega_{Y_p},\cdots 
  \right\} \subset H^4(Y; \C)\, ,
\label{eq:Hscan-min}
\end{equation}
which does not depend on the choice of $p \in {\cal M}_*$.

In order to see how $H_{\rm scan}$ should be chosen to achieve the goal we 
have set in this article, note that the vector space $H^4(Y; \R)$ is decomposed 
as follows:
\begin{equation}
H^4_{H*}(Y; \R) \oplus H^{2,2}_{RM*}(Y; \R) \oplus  H^{2,2}_{V*}(Y; \R)\, ;
\end{equation}
the vertical component $H^{2,2}_{V*}(Y; \R)$ 
is the subspace generated by the wedge products of integral $(1,1)$-forms 
on $Y_p$, which defines a subspace of $H^4(Y;\R)$ independent of 
$p \in {\cal M}_*$. The remaining component $H^{2,2}_{RM*}(Y;\R)$, thus, 
should not depend on $p \in {\cal M}_*$.

We choose $H_{\rm scan}\otimes \R$ to be $H^4_{H*}(Y;\R)$ in this article. 
As discussed in more detail in \cite{BW-H22}, $H^4_{H*}(Y; \R)$ is 
contained in $[H^4(Y;\R)]_{\rm ker}$. Thus, for this choice of $H_{\rm scan}$, 
all the vacua share the same symmetry group from 7-branes in the 
effective theories below the Kaluza--Klein scale. 
This argument does not exclude the option to take $H_{\rm scan} \otimes \R$ 
larger that $H^4_{H*}(Y; \R)$, but we should not take it to be as large 
as $[H^{2,2}_{V*}(Y;\R)]^\perp$. It is not hard to find families
$\pi: {\cal Y} \longrightarrow {\cal M}_*$ with $h^{2,0}(Y_p)=0$ 
where there are algebraic four-forms in $H^{2,2}_{RM*}(Y;\R) \cap H^4(Y;\Z)$ 
that break the symmetry of the 7-brane gauge group \cite{BW-H22}.
A similar phenomenon has also been observed in \cite{BKW-phys}, 
where $Y={\rm K3} \times {\rm K3}$, and the four-form in 
$H^{2,2}_{RM*}(Y;\R)$ is not dual to an algebraic cycle.
\\
\\

\noindent
{\bf 3. {\em The Results}:}
\label{sec:Results}

Let us apply the method described in the previous section to 
derive statistical distributions of observables of practical interest. 
The study in \cite{BKW-phys} used a family of ${\rm K3} \times {\rm K3}$ 
for compactification of F-theory, where all the 7-branes are parallel, and 
there is no light matter fields except those in the adjoint representation 
of the 7-brane gauge groups. Ensembles of flux vacua in such a set-up do 
not include low-energy effective theories that look close to the 
(supersymmetric extensions of the) Standard Model. In this article, we 
therefore use a few other families of elliptically fibred Calabi--Yau fourfolds,
for which the low-energy effective theories are at least semi-realistic. 
These effective theories have $\SU(5)$ (resp. $\SO(10)$) unification, and $N_{\rm gen}$ generations 
of matter fields in the ${\bf 10}+\bar{\bf 5}$ (resp. ${\bf 16}$) 
representations.  The task is to determine the value of $L_*$ and $K$ for 
such families and to study how those values depend on the choice of the 
unification group or the number of generations $N_{\rm gen}$.

\vspace{10pt}
\noindent
{\bf 3.A {\em Number of Generations}}

We focus on a few choices of $(B_3, [S])$ for which 
the restricted moduli spaces and families for SU(5) unification are constructed 
in the way stated at the beginning of the previous section as examples.   
We choose 
\begin{equation}
  B_3 = \mathbb{P} \left[ 
     {\cal O}_{\mathbb{P}^2} \oplus {\cal O}_{\mathbb{P}^2}(n) \right],
 \qquad -3 \leq n \leq 3, 
\label{eq:B3-def}
\end{equation}
which is a $\mathbb{P}^1$-fibration over $\mathbb{P}^2$, 
and let $[S]$ be the ``north-pole section'' of the $\mathbb{P}^1$-fibration
corresponding to the zero of a section of ${\cal O}_{\mathbb{P}^2}$.
The range of $n$ is set so that the 7-brane unification group at 
the $S$ can be as small as ${\rm SU}(5)$, while the gauge group at the 
hidden sector (corresponding to the zero of a section of 
${\cal O}_{\mathbb{P}^2}(n)$) can be completely Higgsed away.

We set $(H_{\rm scan}\otimes \R) = H^4_{H*}(Y;\R)$, and choose 
$G^{(4)}_{\rm fix}$ to be the F-theory dual of the chirality-generating 
bundle twist in \cite{FMW}, parametrized by $\lambda_{FMW} \in 1/2 + \Z$. 
This flux gives rise to the net chirality 
of the matter fields in the ${\bf 10}$ vs $\overline{\bf 10}$
(also $\bar{\bf 5}$ vs ${\bf 5}$) representation of the ${\rm SU}(5)$ 
unification group; $N_{\rm gen} = - (18-n)(3-n) \lambda_{FMW}$ \cite{C-DI}. 
Since the vanishing cycles for the chiral matter belong to $H^{2,2}_{V*}(Y;\Q)$, 
any flux vacua in the ensemble (\ref{eq:flux-ensemble}) have the same 
$N_{\rm gen}$ \cite{BW-H22}. 
In this way, we obtain an ensemble of F-theory flux vacua that share 
the same algebraic and topological ($N_{\rm gen}$) information.
The F-theory dual description of $G^{(4)}_{\rm fix}$ has been determined in \cite{GH}.

$L_* = \chi(Y)/24 - (G^{(4)}_{\rm fix})^2/2$ is the upper bound 
on the net D3-brane charges from $G_{\rm scan}^{(4)}$. It depends 
on $N_{\rm gen}$ through $G^{(4)}_{\rm fix}$ \cite{BKW-phys} and 
a straightforward computation reveals that \cite{GH, BW-H22}
\begin{align}
L_* =  \frac{2163}{4} + \frac{125}{8}\,n\,(n+7)
- \frac{5 \,N_{\rm gen}^2}{2\,(18-n)(3-n)} \, .
\label{eq:tadpole-Gchi}
\end{align}
The maximal values of $L_*^{\rm max}$ range within $\sim 300$ to $800$ for 
the families with $-3 \leq n \leq 3$ (Table \ref{tab:Hodge-nmbr-SU(5)});  
details of the calculation are found in \cite{BW-H22}.
\begin{table}[tbp]
\begin{center}
\caption{\label{tab:Hodge-nmbr-SU(5)} 
Various topological data of the families 
${\cal Y} \rightarrow {\cal M}^{A4}_*$ of Calabi--Yau fourfolds 
for SU(5) unification, with $B_3$ given in (\ref{eq:B3-def}). 
For $n=3$, $L_*^{\rm max}=\chi(Y)/24$, as $G^{(4)}_{\rm fix}=N_{\rm gen}=0$.
} 
\begin{tabular}{|c||ccccccc|}
\hline 
$n$ & $-3$ & $-2$ & $-1$ & 0 & 1 & 2 & 3 \\
\hline 
$h^{3,1}_*$ & 1249 & 1423 & 1723 & 2148 & 2698 & 3373 & 4173 \\
$h^{2,2}_{H*}$ & 5057 & 5755 & 6955 & 8655 & 10855 & 13555 & 16756 \\
$h^{2,2}_{V*}$ & 9 & 9 & 9 & 9 & 9 & 9 & 8 \\  
$L_*^{\rm max}$ & 237 & 297 & 387 & 507 & 657 & 837 & [1047] \\
$K$ & 7557 & 8603 & 10403 & 12953 & 16253 & 20303 & 25104 \\
\hline 
\end{tabular}
\end{center}
\end{table}
$L_*$ is always an upper convex quadratic function of $N_{\rm gen}$ 
in F-theory, not just for the choice in (\ref{eq:B3-def}). 

The other number we need to use in (\ref{eq:ADD-formula-1}) is
$K={\rm dim}_\R H^4_{H*}(Y;\R)$. This task boils down 
to the determination of the dimension of the horizontal component 
$H^{2,2}_{H*}(Y_p; \R)$, since $K = 2 + 2m + {\rm dim}_\R[H^{2,2}_{H*}(Y_p;\R)]$.
A general recipe is to use mirror symmetry and determine 
the dimension of the vertical component of the mirror 
manifold of $Y$. 
The authors derived in \cite{BW-H22} the formula for $h^{2,2}_V$, $h^{2,2}_H$ 
and $h^{2,2}_{RM}$ that is valid for any Calabi--Yau hypersurface of a toric 
5-dimensional ambient space. 

We carried out the computation of $h^{2,2}_{V*}(Y)$, 
$h^{2,2}_{H*}(Y)$ and $K$; for the families for ${\rm SU}(5)$ unification 
with $B_3$ in (\ref{eq:B3-def}), it turns out that $h^{2,2}_{RM}=0$.
Details of the computation 
are found in \cite{BW-H22}, and only the results are recorded in 
Table \ref{tab:Hodge-nmbr-SU(5)}. Certainly the results on 
$L_*^{\rm max}$ and $K$ in this table are only for a limited number of 
choices of $B_3$ and do not tell us whether they are typical among the results 
for all other choices of $B_3$ from 3-dimensional Fano varieties, or 
how much the values of $L_*^{\rm max}$ and $K$ can vary for different $B_3$.
But the table at least provides the first example of such calculations. 
 
With this preparation, we can derive the distribution 
of the number of generations $N_{\rm gen}$ almost immediately. 
We have constructed ensembles 
of flux vacua that are labelled by $\lambda_{FMW} \in 1/2 + \Z$. Each one 
of those ensembles consists of vacua that lead to effective theories 
with common algebraic information (SU(5) GUT with chiral matter in the 
${\bf 10}+\bar{\bf 5}$ representation), but their topological information 
$N_{\rm gen} \propto \lambda_{FMW}$ varies from one ensemble to another. 
To compare the number of vacua that the individual ensembles contain, 
one simply needs to integrate $\rho_I$ over ${\cal M}_*$, which is to 
ignore the difference in the value of the effective coupling constants of 
the vacua in a given ensemble. Since the integral of $\rho_I$ over 
${\cal M}_*$ usually yields a number of order unity (some region of 
${\cal M}_*$ may have to be excluded; cf the discussion of D-limits in \cite{ADD}), 
we simply make an approximation $\int_{{\cal M}_*} \rho_I \approx 1$. 
Only the prefactor $(2\pi L_*)^{K/2}/[(K/2)!]$ in (\ref{eq:ADD-formula-1}) 
is then used as an estimate of the number of vacua in a given ensemble. 
We can use the value of $K$ in Table \ref{tab:Hodge-nmbr-SU(5)}, 
and the $N_{\rm gen}$-dependence of $L_*$ has already been discussed 
in this article. 
The number of vacua depends on $N_{\rm gen}$ in a way the volume of a
$K$-dimensional sphere changes as the radius-square $L_*$ decreases 
quadratically in $N_{\rm gen}$. There is an absolute upper limit on 
$N_{\rm gen}$ for a given family $\pi:{\cal Y} \longrightarrow {\cal M}_*^{A4}$ 
due to the D3-tadpole constraint $L_* \geq 0$.

In the examples of $(B_3, [S])$ we have chosen at the beginning of this 
section, however, we have $L_* \ll K$ (see Table \ref{tab:Hodge-nmbr-SU(5)}). 
The continuous approximation \cite{ADD} is not particularly good 
for such families, and should be replaced 
by $e^{L_* \ln\left(\sqrt{2\pi} K/L_*\right)}$, see \cite{BW-H22} for a more 
detailed discussion.


In fact, it is commonplace to find $L_* \ll K$, not just in 
the familes cosidered in this section, but in a broader class of familes 
$\pi: {\cal Y} \longrightarrow {\cal M}_*^R$ of interest. For $B_3$ 
with $h^{1,1}(B_3) \approx {\cal O}(1)$, for example, $h^{1,1}(Y)$ still 
remains ${\cal O}(1)$, while $h^{3,1}$ is much larger, due to the 
degrees of freedom for the complex structure of fibration. It is then 
a quite natural consequence that $h^{2,2}_V$ and $h^{2,2}_{RM}$ are not 
as large as $h^{2,2}_H$. From this, we find that 
$K \approx \chi(Y) \approx 24 L_*^{\rm max} \approx 8 \pi L_*^{\rm max} \gg L_*$. 
The number of flux vacua for a family ${\cal Y} \rightarrow {\cal M}_*^R$
is then estimated by 
\begin{equation}
e^{L_* \ln\left(\sqrt{2\pi} K/L_*\right)} \approx e^{K/6} e^{- 5 c N_{\rm gen}^2},
\label{eq:distrib-factor}
\end{equation}
where $c$ is the coefficient of $N_{\rm gen}^2$ in (\ref{eq:tadpole-Gchi}), which 
remains of order unity. The expansion in $N_{\rm gen}^2$ in the exponent is 
valid for $N_{\rm gen} \lesssim 10$, since $\chi(Y)/24$ is often around 
$100$--$1000$. The first factor of (\ref{eq:distrib-factor}) depends 
on the choice of the 7-brane symmetry $R$ (and on $(B_3, [S])$), while 
the second factor is a Gaussian distribution on $N_{\rm gen}$ for robust choice 
of $(B_3, [S])$. Algebraic and topological data of effective theories have 
a factorized distribution. One may further bring the distribution 
$\rho_I$ of the effective coupling constants back to (\ref{eq:distrib-factor}), without integrating it over ${\cal M}_*^R$.

\vspace{10pt}
\noindent
{\bf 3.B {\em Cost of Higher-Rank Gauge Groups}}
\\
The distribution (\ref{eq:ADD-formula-1}, \ref{eq:ADD-formula-2}, 
\ref{eq:distrib-factor}) can be used to derive the statistical cost 
of requiring a higher rank gauge group on 7-branes. This idea was pursued 
already in \cite{BKW-phys}, using $Y={\rm K3} \times {\rm K3}$ for 
F-theory compactification; more examples are obtained in this article to 
estimate the systematics. The choices of $(B_3,[S])$ here are also more 
realistic than that of K3 x K3. 

\begin{table}[tbp]
\begin{center}
\caption{\label{tab:compare-diff-rank}
Data for families over ${\cal M}$, ${\cal M}_*^{A4}$ and 
${\cal M}_*^{D5}$.}
\begin{tabular}{||c||c|c|c|c|c|c||}
\hline 
 & $h^{1,1}$ & $h^{3,1}$ & $h^{2,2}_{V*}$ & $h^{2,2}_{H*}$ & 
  $\chi(Y)$ & $K$ \\
\hline 
no gauge group & 3 & 3277 & 4 & 13160 & 19728 & 19716 \\ 
SU(5) model & 7 & 2148 & 9 & 8655 & 12978 & 12953 \\
SO(10) model & 8 & 2138 & 10 & 8618 & 12924 & 12896 \\
\hline 
\end{tabular}
\end{center}
\end{table}
The SO(10) version of the family, 
$\pi: {\cal Y} \longrightarrow {\cal M}_*^{D5}$, 
can also be constructed for the choice of $(B_3, [S])$ 
in (\ref{eq:B3-def}) using toric geometry. 
Various topological data for the families over ${\cal M}_*^{D5}$, 
${\cal M}_*^{A4}$ and ${\cal M}$ (where no 7-brane gauge symmetry is required) 
can be computed and the results are recorded in 
Table \ref{tab:compare-diff-rank} for the $n=0$ case. 
For more details of computations, see \cite{BW-H22}.

One can see from (\ref{eq:distrib-factor}) that the difference in the 
value of $K$ for different choices of 7-brane symmetry $R$ determines 
the relative number of the corresponding flux quanta (vacua).
Ensembles with higher rank 7-brane gauge group have smaller dimension $K$  
(Table~\ref{tab:compare-diff-rank}), confirming the same observation 
in \cite{BKW-phys} based on the family $Y = {\rm K3} \times {\rm K3}$. 
This leads to the observation that the rank-4 gauge group on 7-branes is 
{\it not} as statistically ``natural'' as vacua without a gauge group 
on 7-branes. In the choice of $(B_3,[S])$ in (\ref{eq:B3-def}) with $n=0$, 
for example, vacua with the rank-4 SU(5) unification constitutes only 
the fraction $e^{-\Delta K/6} \approx e^{-1000}$ of the entire flux vacua
(smaller than the fraction $10^{-120}$ for the cosmological constant). 
The authors do not provide their interpretations for this inconvenient 
prediction; a popular attitude will be to hint at poor understanding 
of string theory, to count on cosmological factors that we did not study here, 
and/or to resort to anthropics.
\\
\\
\noindent
{\bf 4. {\em Outlook}:}
\\
This article only deals with the easiest applications of the method 
explained in section 2.
Various ideas of using (\ref{eq:distrib-factor}) and $\rho_I$ 
to address questions of practical interest are described 
in detail in \cite{BW-H22}.
\\
\\
\noindent
{\bf\em Acknowledgements:}
This work is supported in part by STFC under grant ST/J002798/1 (APB), and 
WPI program and Grant-in-Aid for Innovative Areas 2303 from MEXT, Japan (TW).
\vspace{-.6cm}

\end{document}